\documentclass[aps,prl,twocolumn,superscriptaddress,showpacs]{revtex4-1}
\usepackage{graphicx}
\pdfpageattr{/Group <</S /Transparency /I true /CS /DeviceRGB>>}
\usepackage{amsmath,amsthm,amssymb}

\usepackage{color}

\begin{document}

% Use the \preprint command to place your local institutional report
% number in the upper righthand corner of the title page in preprint mode.
% Multiple \preprint commands are allowed.
% Use the 'preprintnumbers' class option to override journal defaults
% to display numbers if necessary
%\preprint{}

%Title of paper
%\title{Confinement-Stabilized Vortex State of Bacterial Suspensions}
%\title{Vortex State of Bacterial Suspensions Stabilized by Confinement} %REG
%\title{Stabilization of Isolated Bacterial Vortices by Confinement} %FGW
%\title{Confinement Stabilizes a Spiral Vortex State of Bacterial Suspensions}
\title{Confinement Stabilizes a Bacterial Suspension into a Spiral Vortex}

\author{Hugo Wioland}
%\email[]{Your e-mail address}
%\homepage[]{Your web page}
%\thanks{}
%\altaffiliation{}
\affiliation{Department of Applied Mathematics and Theoretical Physics, Centre for Mathematical
Sciences, University of Cambridge, Wilberforce Road, Cambridge CB3 0WA, UK}
\author{Francis G. Woodhouse}
\affiliation{Department of Applied Mathematics and Theoretical Physics, Centre for Mathematical
Sciences, University of Cambridge, Wilberforce Road, Cambridge CB3 0WA, UK}
\author{J\"orn Dunkel}
\affiliation{Department of Applied Mathematics and Theoretical Physics, Centre for Mathematical
Sciences, University of Cambridge, Wilberforce Road, Cambridge CB3 0WA, UK}
\author{John O. Kessler}
\affiliation{Department of Physics, University of Arizona, Tucson, AZ 85721, USA}
\author{Raymond E. Goldstein}
\affiliation{Department of Applied Mathematics and Theoretical Physics, Centre for Mathematical
Sciences, University of Cambridge, Wilberforce Road, Cambridge CB3 0WA, UK}

%Collaboration name if desired (requires use of superscriptaddress
%option in \documentclass). \noaffiliation is required (may also be
%used with the \author command).
%\collaboration can be followed by \email, \homepage, \thanks as well.
%\collaboration{}
%\noaffiliation

\date{\today}

\begin{abstract}
Confining surfaces play crucial roles in dynamics, transport and order in many physical systems,
but their effects on active matter, a broad class of dynamically self-organizing systems, are poorly understood. We investigate here 
the influence of global confinement and surface curvature on collective motion by studying the flow and orientational order within 
small droplets of a dense bacterial suspension.
The competition between radial confinement, self-propulsion, steric interactions and hydrodynamics robustly 
induces an intriguing steady single-vortex state, in which cells align in inwardly-spiralling patterns accompanied by a thin 
counterrotating boundary layer. A minimal continuum model is shown to be 
in good agreement with these observations.
\end{abstract}

%\pacs{87.19.Gh}


%47.63.-b biological flow
%47.54.-r pattern selection fluid mechanics
%87.	Biological and medical physics
% 87.10.-e	General theory and mathematical aspects
% 87.10.Ed	Ordinary differential equations (ODE), partial differential equations (PDE), integrodifferential models
%87.18.Hf	Spatiotemporal pattern formation in cellular populations

\maketitle

%\paragraph{Introduction}

Geometric boundaries and surface interactions are known to have profound effects 
on transport and order in condensed matter systems, with examples ranging from nanoscale edge currents in quantum 
Hall devices~\cite{vonklitzing1993,macdonald1984} to macroscopic topological frustration in liquid crystals (LCs) tuned by 
manipulating molecular alignment at confining surfaces~\cite{sec2012}.  By contrast, in 
spite of considerable recent interest~\cite{grossman2008, wensink2008,woodhouse2012,ravnik2013,brotto2013}, 
the effects of external geometric constraints and confining interfaces on collective dynamics of \textit{active}
biological matter~\cite{vicsek1995,koch2011}, such as polar gels~\cite{furthauer2012,kruse2004} 
and bacterial~\cite{KandW,dombrowski2004, tuval2005,cisneros2007,sokolov2007,sokolov2012} or algal 
suspensions~\cite{rafai2010}, are not yet well understood, 
not least owing to a lack of well-controlled experimental systems.

At high concentrations, motile rod-like cells exhibit self-organization akin to nematic LC 
ordering~\cite{KandW,dombrowski2004,tsimring2008}, 
with the added facet of polar alignment driven by collective swimming \cite{ginelli2010, wensink2012b}.  Unlike passive LCs, 
cellular suspensions are in a constant state of flux:  at scales between 10~$\mu\text{m}$ and 1~mm, 
coherent structures (swirls, jets, and plumes) continually emerge and persist for 
seconds at a time~\cite{dombrowski2004, tuval2005, cisneros2007, sokolov2007, wensink2012}.  
While the dynamics of dense bacterial suspensions {\it in bulk} are fairly well
understood~\cite{cisneros2007,sokolov2012,wensink2012, dunkel2013,lushi2013}, microorganisms often live in porous habitats 
like soil, where encounters with interfaces or three-phase contact lines are common~\cite{KandW,dombrowski2004,durham2012}. Recent work has clarified 
how single cells interact with surfaces~\cite{berg2005,tang2008,drescher2011, kantsler2013}, but it 
remains unclear how global geometric constraints influence their collective motion. 

\begin{figure}[b]
	\includegraphics[width = \columnwidth]{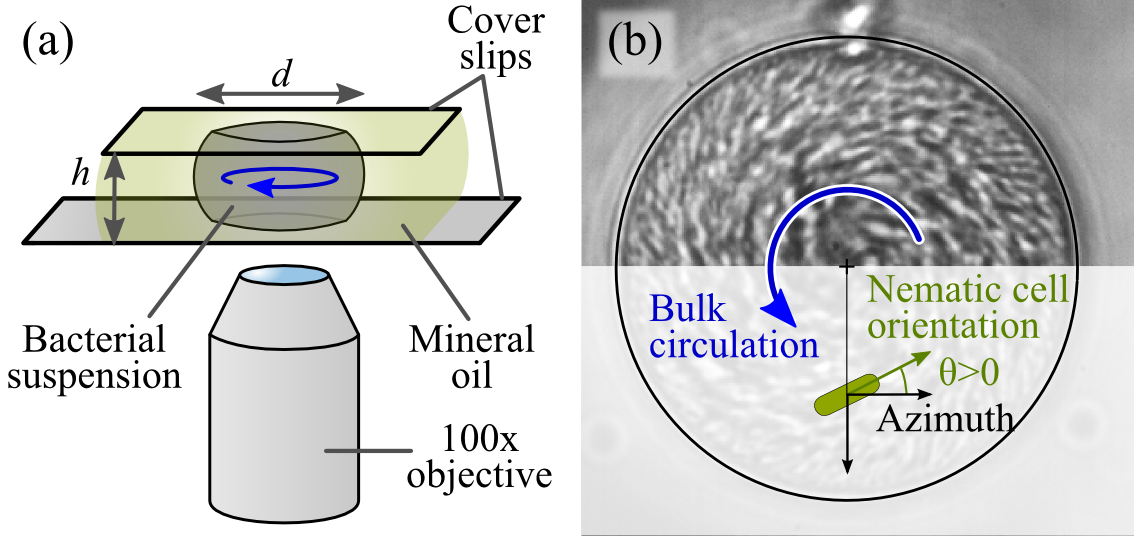}
	\caption{\label{setup} (color online).  Overview. (a) Experimental setup. (b) Bright field image of a $40\,\mu$m drop, 
and definition of cell orientation angle relative to main circulation direction.}
\end{figure}

Here we combine experiment and theory to investigate how confinement and boundary curvature affect stability 
and topology of collective dynamics in active suspensions. The physical system we study is an oil emulsion containing 
droplets of a highly concentrated aqueous suspension of \textit{Bacillus subtilis} (Fig.~\ref{setup}a). For drops 
of diameter~\mbox{$d= \text{30--70}\,\mu$m} and height $h\sim 25\,\mu$m, we find that the suspension
 self-organizes into a single stable vortex~(Fig.~\ref{setup}b)  that persists as long as oxygen is available.  
This pattern is reminiscent of structures seen in colonies on the surface of agar \cite{czirok1996}, 
spontaneously circulating cytoplasmic extracts of algal cells~\cite{woodhouse2012}, 
and the rotating interior of fibroblasts on micropatterned surfaces~\cite{Kumar}.
The vortex flow described here is purely azimuthal and accompanied by a thin counterrotating boundary layer, consisting 
of cells swimming opposite to the bulk. Surprisingly, we observe that the cells arrange in 
spirals with a maximum pitch angle of up to $35^{\circ}$ relative to the azimuthal bulk flow direction~(Fig.~\ref{setup}b). 
We suggest that this intriguing helical pattern results from the interplay of boundary curvature and steric and 
hydrodynamic interactions.   Building on this hypothesis, we formulate a simple continuum model and find good 
agreement between its predictions and experimental results.

%%%%%%%%%%%%
%%%%%%%%%%%%

%%%%%%%%%%%%
\begin{figure*}[t]
	\includegraphics[width = \textwidth]{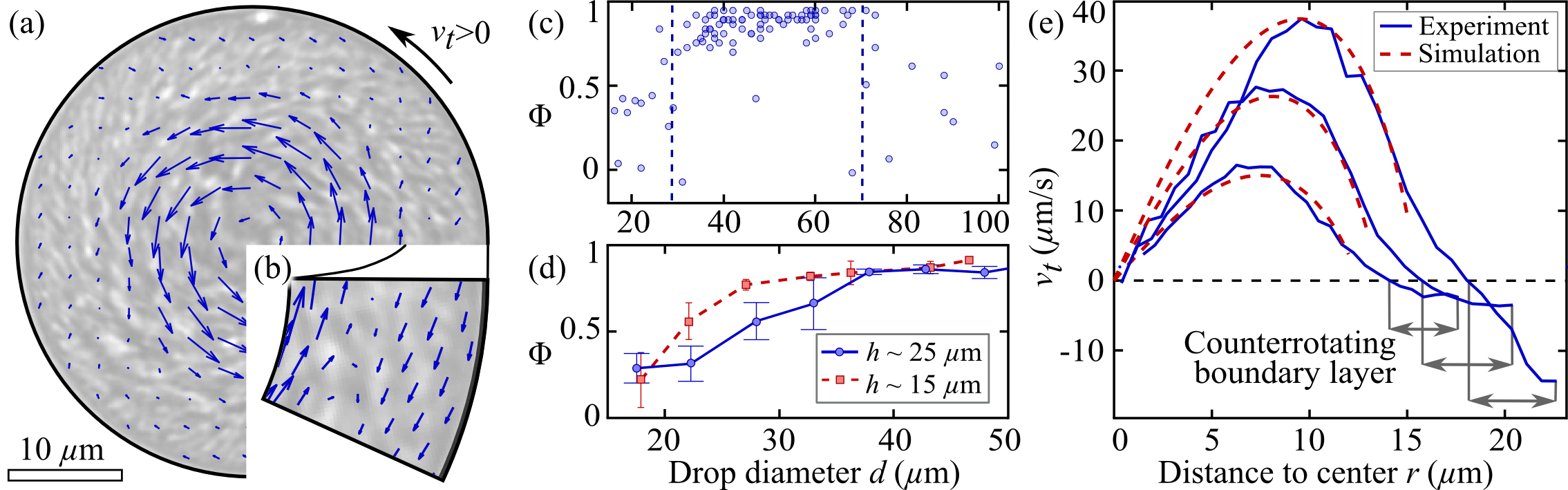}
	\caption{\label{flow} (color online).  Steady-state circulation in highly concentrated {\it B.~subtilis} 
droplet. (a) PIV flow field for a droplet with a volume filling fraction~$\varphi\sim 0.4$. For clarity, not all PIV 
vectors are shown. (b) Enlarged region reveals the counterrotating boundary layer. All PIV 
vectors are shown. (c-d) Vortex order parameter~$V$ for varying diameter~$d$. (c) Drops of constant height 
$h\sim25\,\mu$m. Dashed lines denote the highly ordered single-vortex regime. (d) Averaged vortex order 
parameter~$\Phi$ ($5\,\mu$m bins) for $h\sim15\,\mu$m (red dashed line) and $h\sim25\,\mu$m (blue full line). 
Error bars indicate the standard deviation. (e) Azimuthal flow~$v_t(r) = \langle \mathbf{v}\cdot\mathbf{t} \rangle_\theta$
profile for three different experiments (blue full lines), compared with continuum bulk flow model results (red 
dashed lines). Negative flow indicates the counterrotating boundary layer.\\}
\end{figure*}
%%%%%%%%%%%%

\textit{B. subtilis} (wild-type strain 168) were grown in standard Terrific Broth (TB, Sigma) at $35^{\circ}\,$C on a shaker.  
An overnight culture was diluted 200$\times$ and grown for 5$\,$h until the end of exponential growth when the 
proportion of motile cells is maximal~\cite{kearns2005}. Cells were then centrifuged at $1500g$ for $10\,$min. 
The pellet was gently mixed and transferred to 4 volumes of mineral oil, with $10\,$mg/mL diphytanoyl 
phosphatidylcholine  (DiPhyPC, Avanti) added to prevent the emulsion from coalescing. Small drops were 
created by slowly pipetting the suspension, $10\,\mu$L of which was placed between two coverslips such that 
it spread by surface tension to the coverslip edge. This procedure yields many flattened 
drops with $h\sim25\,\mu$m and diameters ranging from $\text{10--150}\,\mu$m, and bacterial volume 
fraction $\varphi\sim0.4$. Bacteria remain active for several minutes in the largest drops and up to 
$20$ minutes for the smallest, reflecting the larger diffusive influx of oxygen in the smaller drops. Coverslips were rendered hydrophobic with silane, resulting in pancake-shaped drops that 
are wider at the midplane of the chamber than at the top and bottom (Fig.~\ref{setup}a).
Movies were acquired at $125$ fps with a high-speed camera (Fastcam, Photron) on an inverted 
microscope (Cell Observer, Zeiss), using a 100$\times$ oil-immersion objective and analyzed with custom 
Matlab algorithms.  Flows were imaged in the 
center of the chamber to minimize optical distortions.

Confinement by the oil interface stabilizes rapidly rotating vortices (Fig.~\ref{flow} and Supplemental Video 1). 
To quantify this effect, we determined the local bacterial velocity field $\mathbf v(\mathbf{x})$, using a customized 
version of the particle image velocimetry (PIV) toolbox mPIV \cite{mori2006} that averages pixel correlations over 
two seconds~\cite{meinhart2010}. The PIV algorithm yields the local mean velocity of the bacteria, reflecting the 
locomotion due to swimming and advection by the fluid flow  (Fig.~\ref{flow}a). The emergence of stable azimuthal 
flow is captured by the vortex order parameter
\begin{equation}
\Phi = \frac{ { \sum_i{|\mathbf{v}_i\cdot\mathbf{t}_i|} }/{\sum_j{||\mathbf{v}_j||}}-2/\pi}{1-2/\pi},
\end{equation}
where $\mathbf v_i$ is the in-plane velocity and $\mathbf t_i$ the azimuthal unit vector (Fig.~\ref{setup}b) at PIV grid 
point $\mathbf{x}_i$.  $\Phi=1$ for steady azimuthal circulation, $\Phi=0$ for disordered chaotic
 flows and $\Phi<0$ for predominantly radial flows. Plotting $\Phi$ as a function of drop diameter reveals that a 
highly-ordered single-vortex state with $\Phi>0.7$ forms if $d_- <d<d_+$ with $d_-\sim30\,\mu$m and
$d_+\sim70\,\mu$m (Fig.~\ref{flow}c). Clockwise and counterclockwise vortices occur with equal probability. 
The lower critical diameter $d_-$ depends on the chamber height $h$ (Fig. \ref{flow}d). Lowering~$h$ restores the 
quasi-2D nature of the confinement and allows for formation of vortex states at smaller diameter~$d$. The upper 
critical diameter $d_+$ is consistent with the size of the transient turbulent swirls observed in 3D bulk 
bacterial suspensions~\cite{cisneros2007,sokolov2012, dunkel2013}. % but considerably larger 
%than for 2D suspensions~\cite{wensink2012}. 
%For drops smaller than $30\, \mu$m, because the chamber is $25\mu m$ height, the quasi-2D confinement 
%is lost and no steady flow appears in the imaging plane; reducing the chamber height restores the confinement 
%and the minimal diameter for steady circulation decreases.  \textcolor{red}{?? this should be illustrated with data 
% in Fig.2b??}. 
In drops slightly larger than $d_+$ flow is still azimuthal near the boundary regions but the vortex order decreases 
toward the center. Drops with $d\gtrsim 100\, \mu$m show fully developed bacterial turbulence as seen in 
quasi-infinite suspensions~\cite{dombrowski2004,cisneros2007,sokolov2012,dunkel2013}. 
%We observed that vortex structures within such drops have a characteristic diameter of approximately $35 \mu m$. 

%Without lateral confinement, the typical swirl diameter \cite{cisneros2007} and velocity correlation length 
%\cite{wensink2012, dunkel2013} for {\it B. subtilis} is about $30 \mu m$, less than half the size of the largest 
%stable single-vortex states observed here. This remarkable effect is a consequence of the total confinement 
%effectively `quantizing' the allowable states \cite{woodhouse2012}, with vortices that are `too large' dynamically 
%preferred to those that are `too small' compared to the unconfined vortex scale. 

The azimuthal flow speed in a vortex state is maximal at a distance $\sim\! d/4$ from the center (Fig.~\ref{flow}e).
Across experiments, the maximum speed increases with $d$, reaching $\sim\! 40 \mu$m/s for  $d_+$,
roughly four times the typical swimming speed of an isolated bacterium~\cite{sokolov2007} and in agreement with 
measurements in open {\it B. subtilis} suspensions~\cite{sokolov2007, cisneros2007}. 
While our setup does not supply oxygen, and the bacterial motility decreases~\cite{sokolov2012} with time,  
recent studies of quasi-infinite suspensions~\cite{sokolov2012,dunkel2013} have shown that 
the flow correlation length is independent of swimming speed at high cell density, so we may 
neglect oxygen depletion in the analysis of patterns. In the following, we focus on the properties of single-vortex 
states with $\Phi>0.7$ and take the azimuthal unit vector $\mathbf t$ to point in the direction of bulk flow, so that we 
may treat clockwise and counterclockwise vortices equally (Fig.~\ref{setup}b).

Detailed flow field analysis reveals that highly ordered vortex states are always accompanied by a thin layer of cells 
swimming along the oil interface in the opposite direction to the bulk flow (Fig.~\ref{flow}b). This surprising fact is 
reflected in the azimuthally-averaged circulation velocity profile 
$v_t(r) = \langle \mathbf{v}(\mathbf{x})\cdot\mathbf{t} \rangle_\theta$, 
where $\mathbf{x}=(r\cos\theta,r\sin\theta)$, which changes sign towards the edge of the droplet (Fig.~\ref{flow}e). 
The basic form of $v_t(r)$ is preserved among well-ordered droplets ($\Phi>0.7$) with different diameters (Fig. \ref{flow}e). 
To exclude the possibility that the backflow arises from specific interactions between bacteria, DiPhyPC and oil, 
we performed control experiments with dense suspensions in shallow cylindrical polydimethylsiloxane chambers,
and found qualitatively similar behavior.  This result suggests that the formation of a thin counterflow boundary layer is a 
generic phenomenon in bacterial suspensions confined by a higher-viscosity medium.
%%%%%%%%%%%
\begin{figure}[t]
	\includegraphics[width = \columnwidth]{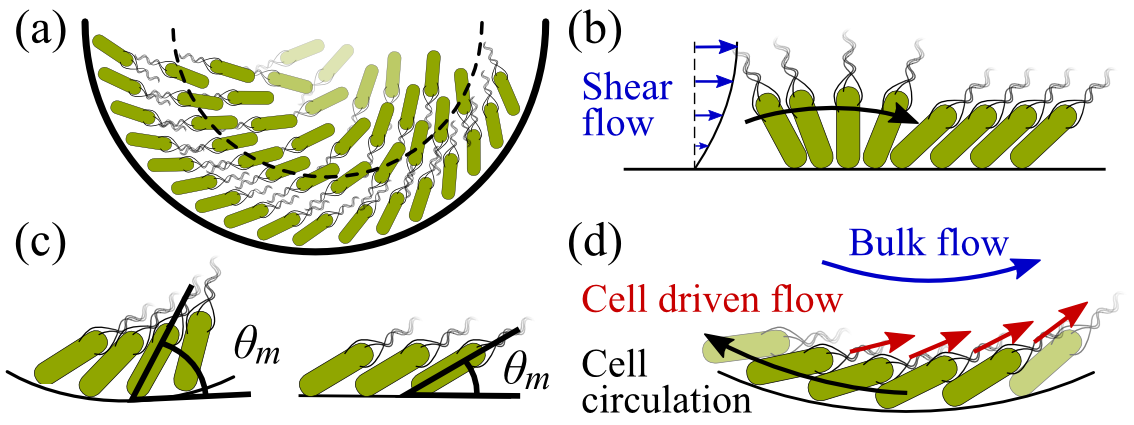}
	\caption{\label{model} (color online).  Schematic cell organization in droplets.  (a) Dashed line 
indicates continuum model boundary, where bulk flow begins. (b-d) Physical mechanisms driving boundary 
layer formation. (b) Shear flow reorients cells to face upstream. (c) Contact angle $\theta_m$  decreases 
with the drop diameter, restricted by steric interactions. (d) Ratchet-like steric repulsion and inward flow 
(red arrows) created by boundary cells force the next layer to move in the opposite 
azimuthal direction, thereby setting the bulk flow direction.}
\end{figure}
%%%%%%%%%%%
By determining the zeros of $v_t$ for all ordered droplets,  we find that the boundary layer thickness
$b$ is independent of $d$ (Fig.~\ref{flow}e). The average value $\bar{b}\approx 4\, \mu$m is slightly smaller than 
the length $\ell\approx5\,\mu$m of \textit{B. subtilis}~\cite{wensink2012}, suggesting that the counterflow region is 
comprised of a single layer of cells.  We tested this hypothesis by imaging droplets in a plane near the bottom 
cover slip in order to resolve vertical cell layers more easily, and confirmed that cells swimming in the 
direction opposite to the bulk flow are in direct contact with the oil interface (Fig.~\ref{model}a and 
Supplemental Video 2).

The presence of this previously unreported counter-flow layer can be understood by considering the main 
forces that cause reorientation of cells near the boundary.  Since the oil viscosity is ten times that of 
water, the interface acts as a nearly-no-slip boundary for the suspension.  Thus, circular bulk motion 
creates a shear flow that exerts torque on the cells in the boundary layer (Fig.~\ref{model}b). As recently shown for 
dilute suspensions~\cite{hill2007}, bacteria prefer to swim upstream when exposed to such flow gradients, thereby 
favoring the formation of a counterrotating layer. If the concentration of cells is sufficiently high, nematic ordering due to 
steric interactions further stabilizes this layer~\cite{vicsek1995, ginelli2010, wensink2012b}. Once the 
layer has formed, cells  trapped in it form a steric ratchet-like structure and, because they are 
pusher-type swimmers~\cite{drescher2011}, they generate a backflow in the direction opposite to their 
orientation (Fig.~\ref{model}d). Both effects force cells in the second layer to move in the other direction: 
the boundary monolayer stabilizes the bulk flow and \textit{vice versa}.
The absence of such counter-circulation in the free-boundary geometry studied by 
Czirok {\it et al.} \cite{czirok1996} provides further evidence that the backflow 
is a consequence of rigid boundary effects. 

%%%%%%%%%%
\begin{figure}
\includegraphics[width = \columnwidth]{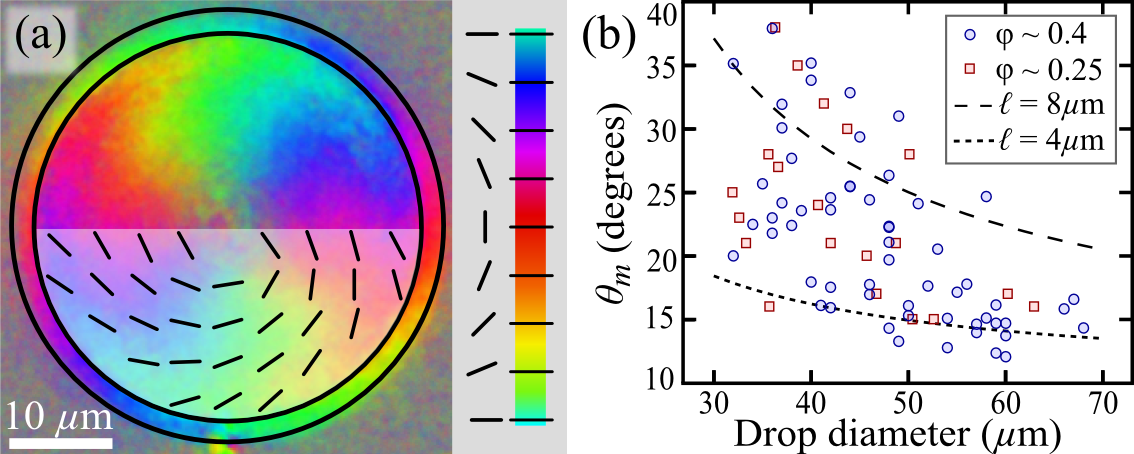}
\caption{\label{orientation} (color online). Bacterial orientation. (a) Local orientation, averaged over $2$ s. 
External ring lies at the water/oil interface and shows local azimuthal direction, and cellular orientation 
appears in the central disc. Discontinuity in color between ring and disc indicates the angle between cells 
and the azimuthal direction. (b) Boundary angle (Fig. \ref{model}c) as a function of drop diameter,
Symbols denote different bacterial concentrations; dashed black lines indicate geometric estimates of 
minimum packing angle $\Theta$ for different cell lengths.}
\end{figure}
%%%%%%%%%%

A dense suspension of rod-like bacteria locally aligns through active nematic interactions 
\cite{vicsek1995, ginelli2010, wensink2012b}. We observe cell orientation that is not parallel to the flow direction: in 
the bulk circulation the cells point inwards, and in the boundary layer they point outwards (Fig.~\ref{model}a).  
We extract the local mean orientation from the bacterial speckle by computing the orientation tensor 
\cite{rezakhaniha2011} (Fig. \ref{orientation}a). As for the flow, we examine the azimuthally-averaged orientation 
angle $\theta(r)$ relative to the circulation direction $\mathbf t$. Near the center of a drop, cells are aligned roughly 
parallel to the bulk circulation ($\theta \approx 0$), and the angle increases with $r$ to a maximum value $\theta_m$ 
close to the boundary. Viewing $\theta_m$ as a function of $d$, we find an inverse correlation: the smaller the drop 
(and thus the higher the boundary curvature), the larger the deviation from the azimuthal direction, 
ranging from $\theta_m\sim10^{\circ}$ for $d=70\,\mu$m to $\theta_m\sim35^{\circ}$ for $d=30\,\mu$m 
(Fig. \ref{orientation}b).
To test whether $\theta_m$ depends on the curvature or on the suspension size, we performed measurements with 
suspensions diluted to $\sim 2/3$ of the starting concentration. In such a drop, cells concentrate at the boundary, 
leaving the center almost empty (Supplemental Video 3). Yet, the measured angles are comparable 
to those of fully concentrated suspensions (Fig. \ref{orientation}b), indicating that this is indeed an effect of 
boundary curvature.

To explain this phenomenon qualitatively, we consider purely steric bacterial packing in the boundary layer. This 
viewpoint is supported by the simulations of Wensink and L{\"o}wen \cite{wensink2008} which show 
that a group of self-propelled particles does not align parallel to a boundary but instead lies at an angle limited by 
steric repulsion. Given a bacterial concentration, we model cells as thin rectangles equally spaced around a circle of 
diameter $d$ and then calculate the minimum packing angle $\Theta$ with the azimuthal direction at which the 
cells could lie in one plane. A dilute suspension thus has edge-parallel packing ($\Theta = 0^{\circ}$), while at some
limiting concentration they become boundary-perpendicular ($\Theta = 90^{\circ}$). In the intermediate regime, 
$\Theta$ decreases as drop diameter $d$ increases (i.e., as boundary curvature falls; Fig. \ref{model}c).
Figure~\ref{orientation}b illustrates packing curves for two cell lengths, $\ell=4\,\mu$m and $\ell=8\,\mu$m, at a 
volume fraction of $0.5$. The measured values of $\theta_m$ then lie between these two curves, indicating that the 
scatter can be explained by variations in the cell length $\ell$ (which are also observed across experiments).

While single-field phenomenological models can describe dense bacterial 
flow quantitatively \cite{wensink2012,dunkel2013}, they do not incorporate the additional observable 
of cellular orientation.  Thus, as is typical in active suspension theory \cite{saintillan2008} we describe the system by 
two functions: the bacterial polar order parameter $\mathbf P$, where $|\mathbf{P}|=0$ for total disorder and 
$|\mathbf{P}|=1$ for total order in direction $\mathbf{P}$, and the suspending fluid flow $\mathbf u$. The fluid 
obeys the forced Stokes equations with friction,
\begin{align*}
-\mu \nabla^2 \mathbf u + \nu \mathbf u + \nabla \Pi = -c_0 \sigma\nabla\cdot(\mathbf{P}\mathbf{P}),
\end{align*}
and incompressibility $\nabla \cdot \mathbf{u} =0$. The viscosity $\mu$ and coefficient of friction $\nu$ 
(from the effects of high bacterial density) control the fluid response to dipolar `pusher' forcing (strength $\sigma$) in
a suspension of concentration $c_0$.  Defining the 
incompressible \emph{swimming field} functional $\mathbf{s}[\mathbf{P}]$, $\nabla \cdot \mathbf{s} = 0$, 
the polar order $\mathbf P$ evolves as
\begin{align*}
\partial_t \mathbf P + (\mathbf u + \mathbf s)\cdot \nabla \mathbf P
&= D_s \nabla^2 \mathbf P - D_r \mathbf P 
+ \alpha(1 - |\mathbf P |^2) \mathbf P \\
&\quad + \epsilon (\mathbb I - \mathbf P  \mathbf P) \cdot (\gamma \mathbf E + \mathbf W) \cdot \mathbf P.
\end{align*}
On the l.h.s, cells are advected by a flow field $\mathbf{u} + \mathbf{s}$, where in 
general, $\mathbf{s}$ is proportional to the incompressible part of $\mathbf{P}$. This ensures that concentration 
fluctuations always dissipate, as appropriate for a highly dense suspension. 
On the r.h.s., the terms are, in order: spatial and rotational diffusion with respective constants $D_s$ and 
$D_r$; spontaneous polar ordering of strength $\alpha$; and reorientation induced by solvent 
strain $\mathbf{E}=(\nabla\mathbf{u} + \nabla\mathbf{u}^\mathsf{T})/2$ and vorticity 
$\mathbf{W}=(\nabla\mathbf{u} - \nabla\mathbf{u}^\mathsf{T})/2$, with cell shape parameter 
$\gamma \in [-1,1]$ and effectiveness $\epsilon \leq 1$ (inhibited by steric effects).
The presence of a bacterial boundary layer in a circular bulk flow is mimicked by 
imposed boundary conditions at $r=d_0/2$ of fixed orientation 
$\mathbf{P} = \mathbf{t}\cos\theta_b - \mathbf{r}\sin\theta_b$, where $\mathbf r$ is the outward radial unit vector. 
A no-slip boundary condition is imposed on the fluid flow.  (A systematic treatment of this model with 
appropriate nondimensionalizations will be presented elsewhere;  
% Note that some of these parameters ($R$, $V$ and $\alpha$, for instance) could be eliminated by non-dimensionalization 
% in a more detailed study of this model. 
here we retain the fully dimensional parameters for simpler connection with 
experiments.)

To model the steady vortex regime we reduce to axisymmetry, where $\mathbf{u} = u \mathbf{t}$ and 
$\mathbf{s} = s \mathbf{t}$ by incompressibility. We then set $s = V\mathbf{P}\cdot\mathbf{t}$ for azimuthal swimming 
at speed $V$. To model the results in Fig.~\ref{flow}d, we fix appropriate parameter values~\cite{drescher2011} 
$c_0 = 0.1\,\mu\text{m}^{-3}$, $\mu = 10^{-3}\,\text{Pa}\,\text{s}$, 
$\nu = 10^{-4}\,\text{Pa}\,\text{s}\,\mu\text{m}^{-2}$, $D_s = 10^3\,\mu\text{m}^2\,\text{s}^{-1}$, 
$D_r = 0.057\,\text{s}^{-1}$, $\alpha = 25\,\text{s}^{-1}$, $\epsilon = 0.5$, $\gamma = 0.9$, and 
$\theta_b = 20^\circ$. We then choose three bulk domain diameters $d_0=24,26,30\,\mu$m, and for each  
we pick $V=4,7,10\,\mu\text{m}\,s^{-1}$ and $\sigma=0.3,0.525,0.75\,\text{pN}\,\mu$m respectively, reflecting 
varying oxygen availability. These yield the steady-state curves of the lab frame bacterial flow 
$|\mathbf{u}+\mathbf{s}|$ shown in Fig.~\ref{flow}c, exhibiting good agreement in the bulk flow regime. 
Additionally, the orientation angle $\theta(r)$ decreases towards zero from its initial
value $\theta(d_0/2) = \theta_b$ as $r$ decreases, as observed experimentally (Fig. \ref{orientation}a).

%We have investigated the collective motion of a dense suspension of rod-like bacteria in total confinement, using small 
%flattened bacterial drops in an oil emulsion. A key finding is that lateral confinement stabilizes the naturally turbulent 
%collective behavior and promotes formation of a single large, steady vortex, whose flow profile can be modeled 
%using a two-field continuum theory. Vortex states consistently exhibited a single layer of counter-circulating cells 
%at the boundary, supplying a boundary condition for a continuum model, and displayed surprisingly large deviations of 
%bulk cellular orientation from the azimuthal direction, reaching deviation angles of $35^{\circ}$.

% FGW: I'm not really sure how this is smectic etc., and the
% Kruse active stuff is basically an active polar LC as well
%The bacterial arrangement observed here is reminiscent of the rotating spirals predicted for totally ordered active gels \cite{kruse2004}, though this model describes actin-myosin cytoskeletal dynamics and lacks interactions particular to microswimmer suspensions \cite{joanny2009}. A better representation could be derived from polar active liquid crystals: the bacterial boundary layer could be regarded as a smectic structure \cite{giomi2008} while the bulk behaves as a chiral nematic phase \cite{sec2012}.

The overall bacterial arrangement we have observed is reminiscent of rotating spirals predicted for totally ordered active 
gels \cite{kruse2004}, although that model describes the actin-myosin cytoskeleton and lacks interactions particular 
to microswimmer suspensions \cite{joanny2009}. A more appropriate representation could be derived from polar active 
liquid crystals: the bacterial boundary layer could be regarded as a smectic structure \cite{giomi2008} while the bulk behaves 
as a chiral nematic phase \cite{sec2012}. 
Yet, it is only by considering the microscopic hydrodynamics near the 
oil interface that the presence of the backflow layer can be inferred. This lends a note of caution to continuum 
modeling of microswimmer suspensions, suggesting that conditions at boundaries, and microscopic effects in general, 
warrant careful and deliberate consideration.  Our combined experimental and theoretical results demonstrate 
that suitably designed boundaries provide a means for stabilizing and controlling order in active microbial systems.

We thank A. Honerkamp-Smith, V. Kantsler, P. Khuc Trong, K. Leptos and E. Lushi for discussions.  This work was supported by the EPSRC and ERC Advanced Investigator Grant 247333.

%old conclusion
%Of course, the experiments presented here are an easily-observed realization of a self-propelled active suspension, a popular topic of current theoretical research. Here, we introduced a simple coupled model of suspension orientation and solvent flow at a minimal level of complexity. It is possible to model flows of concentrated suspensions using a single field theory \cite{wensink2012,dunkel2013}, but our second observable of non-azimuthal bacterial orientation demands a two-field theory here. The overall bacterial arrangement we observed is reminiscent of rotating spirals predicted for totally ordered active gels \cite{kruse2004}. However, this model describes the actin-myosin cytoskeleton and lacks interactions particular to microswimmer suspensions \cite{joanny2009}. A better representation could be derived from polar active liquid crystals: the bacterial boundary layer could be regarded as a smectic structure \cite{giomi2008} while the bulk behaves as a chiral nematic phase \cite{sec2012}. More concretely, though, the bacterial boundary layer is a single-cell effect, providing effective boundary conditions for our continuum model and highlighting the importance of considering potential microscopic effects in naive continuum modeling. {\it Caveat emptor}.

\begin{acknowledgments}
\end{acknowledgments}


\begin{thebibliography}{99}
\bibitem{vonklitzing1993} 
K. von Klitzing, 
Physica B {\bf 184}, 1 (1993).

\bibitem{macdonald1984} 
A. H. MacDonald and P. St\v{r}eda,  
Phys. Rev. B {\bf 29}, 1616 (1984).

\bibitem{sec2012} 
D. Se{\v{c}}, T. Porenta, M. Ravnik, and S. {\v{Z}}umer, 
Soft Matter {\bf 8}, 11982 (2012).

%\bibitem{musevic2006} 
%I. Mu{\v{s}}evi{\v{c}}, M. {\v{S}}karabot, U. Tkalec, M. Ravnik, and S. {\v{Z}}umer, 
%Science {\bf 313}, 954 (2004).

\bibitem{grossman2008} 
D. Grossman, I.S. Aranson, and E.B. Jacob, 
New J. Phys. {\bf 10}, 023036 (2008).

\bibitem{wensink2008} 
H.H. Wensink and H. L\"owen, 
Phys. Rev. E. {\bf 78}, 031409 (2008).

\bibitem{woodhouse2012} 
F.G. Woodhouse and R.E. Goldstein, 
Phys. Rev. Lett. {\bf 109}, 168105 (2012).

\bibitem{ravnik2013} 
M. Ravnik and J.M. Yeomans, 
Phys. Rev. Lett. {\bf 110}, 026001 (2013).

\bibitem{brotto2013} 
T. Brotto, J.-B. Caussin, E. Lauga, and D. Bartolo,
Phys. Rev. Lett. {\bf 110}, 038101 (2013).

\bibitem{koch2011} 
D.L. Koch and G. Subramanian, 
Annu. Rev. Fluid Mech. {\bf 43}, 637 (2011).

\bibitem{vicsek1995} 
T. Vicsek, A. Czir\`ok, E. Ben-Jacob, I. Cohen, and O. Shochet, 
Phys. Rev. Lett. {\bf 75}, 1226 (1995).

\bibitem{furthauer2012} 
S. F\"urthauer, M. Neef, S. W. Grill, K. Kruse, and F. J\"ulicher, 
%The TaylorÐCouette motor: Spontaneous Flows of Active Polar Fluids Between Two Coaxial Cylinders
New J. Phys. {\bf 14}, 023001 (2012).

\bibitem{kruse2004} 
K. Kruse, J.F. Joanny, F. J\"ulicher, J. Prost, and K. Sekimoto, 
Phys. Rev. Lett. {\bf 92}, 078101 (2004).

\bibitem{KandW} J.O. Kessler and M. F. Wojciechowski, in {\it Bacteria as
Multicellular Organisms}, ed. by J. A. Shapiro and M.
Dworkin (Oxford University Press, New York, 1997),
p. 417.

\bibitem{dombrowski2004} 
C. Dombrowski, L. Cisneros, S. Chatkaew, R.E. Goldstein, and J.O. Kessler, 
Phys. Rev. Lett. {\bf 93}, 098103 (2004).

\bibitem{tuval2005} 
I. Tuval, L. Cisneros, C. Dombrowski, C.W. Wolgemuth, J.O. Kessler, and R.E. Goldstein, 
Proc. Natl. Acad. Sci. U.S.A. {\bf 102}, 2277 (2005).

\bibitem{cisneros2007} 
L.H. Cisneros, R. Cortez, C. Dombrowski, R.E. Goldstein, and J.O. Kessler, 
Exp. Fluids {\bf 43}, 737 (2007).

\bibitem{sokolov2007} A. Sokolov, I.S. Aranson, J.O. Kessler, and R.E. Goldstein, Phys. Rev. Lett. {\bf 98}, 158102 (2007).

\bibitem{sokolov2012} 
A. Sokolov and I.S. Aranson, 
Phys. Rev. Lett. {\bf 109}, 248109 (2012).

\bibitem{rafai2010} 
S. Rafai, L. Jibuti, and P. Peyla, 
Phys. Rev. Lett. {\bf 104}, 098102 (2010).

\bibitem{tsimring2008}
D.Volfson, S. Cookson, J. Hasty, and L.S. Tsimring,
%Biomechanical ordering of dense cell populations
Proc. Natl. Acad. Sci. U.S.A. {\bf 105}, 15346 (2008).

\bibitem{ginelli2010} 
F. Ginelli, F. Peruani, M. B\"ar, and H. Chat\'e, 
Phys. Rev. Lett. {\bf 104}, 184502 (2010).

\bibitem{wensink2012b} 
H.H. Wensink, and H. L\"owen, 
J. Phys.: Condens. Matter {\bf 24}, 46413 0(2012).

\bibitem{wensink2012} 
H.H. Wensink, J. Dunkel, S. Heidenreich, K. Drescher, R.E. Goldstein, H. L\"owen, and J.M. Yeomans, 
Proc. Natl. Acad. Sci. U.S.A. {\bf 109}, 14308 (2012).

\bibitem{dunkel2013} 
J. Dunkel, S. Heidenreich, K. Drescher, H.H. Wensink, M. B\"ar, and R.E. Goldstein, 
arXiv:1302.5277 (2013).

\bibitem{lushi2013}
E. Lushi, and C.S. Peskin, Comput. Struct., in press (2013).

\bibitem{durham2012}
W.M. Durham, O. Tranzer, A. Leombruni, R. Stocker,
Phys. Fluids {\bf 24}, 091107 (2012).

\bibitem{berg2005}
W.R. DiLuzio, L. Turner, M. Mayer, P. Garstecki,  D.B. Weibel, H.C. Berg, and G.M. Whitesides,
Nature {\bf 435}, 1271 (2005).

\bibitem{tang2008}
G. Li and L.-K. Tam and J. X. Tang,
Proc. Natl. Acad. Sci. U.S.A. {\bf 105}, 18355 (2008).

\bibitem{kantsler2013} 
V. Kantsler, J. Dunkel, M. Polin, and R.E. Goldstein, 
Proc. Natl. Acad. Sci. U.S.A. {\bf 110}, 1187 (2013).

\bibitem{drescher2011} 
K. Drescher, J. Dunkel, L.H. Cisneros, S. Ganguly, and R.E. Goldstein, 
Proc. Natl. Acad. Sci. U.S.A. {\bf 108}, 10940 (2011).

\bibitem{czirok1996} A. Czir\'ok, E. Ben-Jacob, I. Cohen, and T. Vicsek, 
Phys. Rev. E {\bf 54}, 1791 (1996). 

\bibitem{Kumar} A. Kumar, A. Maitra, M. Sumit, G.V. Shivashankar, and S. Ramaswamy, arxiv:1302.6052.
\bibitem{kearns2005} 
D.B. Kearns and R. Losick, 
Gene Dev. {\bf 19}, 3083 (2005).

\bibitem{mori2006} 
N.~Mori and K.-A.~Chang (2006); For details, see www.oceanwave.jp/softwares/mpiv\_doc.

\bibitem{meinhart2010} 
C.D. Meinhart, S.T. Wereley, and J.G. Santiago, 
J. Fluid. Eng. {\bf 122}, 285 (2000).

\bibitem{hill2007}
J. Hill, O. Kalkanci, J.L. McMurry, and H. Koser, 
Phys. Rev. Lett. {\bf 98}, 068101 (2007).

\bibitem{rezakhaniha2011} 
R. Rezakhaniha {\it et al.},
%A. Agianniotis, J.T.C. Schrauwen, A. Griffa, D. Sage, C.V.C. Bouten, F.N. {van de Vosse}, M. Unser, and N. Stergiopulos, 
Biomech. Model. Mechan. {\bf 11}, 461 (2011).

\bibitem{saintillan2008}
D. Saintillan and M.J. Shelley,
Phys. Rev. Lett. {\bf 100}, 178103 (2008).

\bibitem{joanny2009} J.-F. Joanny and J. Prost, HFSP J. {\bf 3}, 94 (2009).

\bibitem{giomi2008} L. Giomi, M.C. Marchetti, and T.B. Liverpool, Phys. Rev. Lett. {\bf 101}, 198101 (2008).


%\bibitem{sokolov2009} A. Sokolov, R.E. Goldstein, F.I. Feldchtein, and I.S. Aranson, Phys. Rev. E {\bf 80}, 031903 (2009)

%\bibitem{hill2005} N.A. Hill, and T.J. Pedley, Fluid Dyn. Res. {\bf 37}, 1 (2005)



\end{thebibliography}
\end{document}